\documentclass[prl,twocolumn,floatfix,showpacs,superscriptaddress]{revtex4}%
\usepackage{amssymb}
\usepackage{amsmath}
\usepackage{graphicx}
\usepackage{epsfig}
\usepackage{amsfonts}%
\providecommand{\U}[1]{\protect\rule{.1in}{.1in}}
\begin{document}
\title{Generalized Stern-Gerlach Effect for Chiral Molecules}
\author{Yong Li}
\affiliation{Department of Physics, University of Basel, Klingelbergstrasse
82, 4056 Basel, Switzerland}
\author{C. Bruder}
\affiliation{Department of Physics, University of Basel, Klingelbergstrasse
82, 4056 Basel, Switzerland}
\author{C. P. Sun}
\affiliation{Institute of Theoretical Physics, Chinese Academy of Sciences, Beijing,
100080, China}
\date{\today }

\begin{abstract}
The Stern-Gerlach effect is well-known as spin-dependent splitting of a beam
of atoms with magnetic moments by a magnetic-field gradient. Here, we show
that an induced gauge potential may lead to a similar effect for chiral
molecules. In the presence of three inhomogeneous light fields, the
center-of-mass of a three-level chiral molecule is subject to an optically
induced gauge potential, and the internal dynamics of the molecules can be
described as an adiabatic evolution in the reduced pseudo-spin subspace of the
two lowest energy levels. We demonstrate numerically that such an induced
gauge potential can lead to observable pseudo-spin dependent and
chirality-dependent generalized Stern-Gerlach effects for mixed left- and right-handed chiral molecules under realistic conditions.

\end{abstract}

\maketitle


\emph{Introduction.} The Stern-Gerlach experiment \cite{SG} is one of the
milestones in the development of quantum theory. The observation of the
splitting of a beam of silver atoms in their ground states in a non-uniform
magnetic field led to the concept of the electronic spin. In this paper we
will show that even in the absence of a magnetic-field gradient, the
center-of-mass of certain atoms or molecules in optical fields will follow
different trajectories corresponding to different inner states. This
phenomenon is straightforwardly explained as a generalized Stern-Gerlach
effect by the optically-induced gauge potential \cite{wilz,moody,sun-ge}.

This induced gauge potential consists of the effective vector and scalar
potentials, which result from the adiabatic variable separation of the slow
spatial and fast inner dynamics of the atom according to the generalized
Born-Oppenheimer approximation \cite{sun-ge}. Recently, there has been
considerable interest to implement various pseudo-spin dependent induced gauge
potentials for cold atoms. Examples include the induced monopole
\cite{zhang-li-sun}, and the spin Hall effect for cold atoms \cite{zhu,liu},
in direct analogy to the spin Hall effect due to the spin-orbit coupling in
condensed matter physics \cite{SHE}. Here, we would like to consider
consequences of the induced gauge potential in systems of cold chiral
molecules \cite{Kral01,Kral03,shapiro} that manifest themselves as a 
generalized Stern-Gerlach effect.

\begin{figure}[ptbh]
\includegraphics[width=6cm]{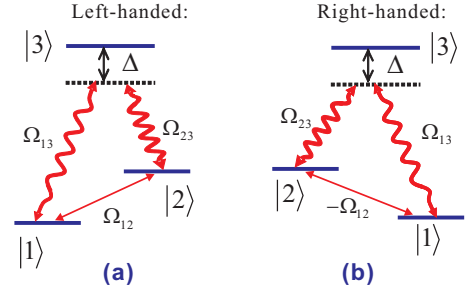}\caption{ (Color online) Model of
three-level $\Delta$-type left-(a) and right-(b) handed chiral molecules,
coupled to laser beams with Rabi frequencies $\pm\Omega_{12}$, $\Omega_{13}$,
and $\Omega_{23} $.}%
\label{Fig01}%
\end{figure}

We consider a chiral molecule (see Fig.~\ref{Fig01}), which is
described by a cyclic three-level system
\cite{Kral01,Kral03,Liu05,cyclic,sun05} where any two of the three
levels are coupled by a classical optical field.  A specific example
are cyclic three-level ($\Delta$-type) chiral molecules, e.g., the
$D_2S_2$ enantiomers in Ref. \cite{Kral03} when only the lowest three
states in each well are considered. Such symmetry-breaking systems can
also be implemented using an asymmetric well and its mirror
\cite{Kral01} (i.e., one asymmetric well and its mirror form a
symmetric double well which supports chirality), or a superconducting
circuit acting as an effective atom \cite{Liu05}. It will be shown
that the optically-induced gauge potentials for the chiral molecules
will be both chirality-dependent and pseudo-spin-dependent when the
internal dynamics of chiral molecules are described as an adiabatic
evolution in the reduced pseudo-spin subspace of the two lowest energy
levels. Thus, the generalized Stern-Gerlach effect can be used to
distinguish molecules with different chiralities, suggesting a
discrimination method to separate chiral mixtures.

\emph{Model.} We first consider a general case of symmetry-breaking molecule
having a $\Delta$-type or cyclic three-level configuration (e.g., see the
left-handed chiral molecule in Fig.~\ref{Fig01}(a)). The ground state
$|1\rangle$ and the metastable state $|2\rangle$ are coupled to the excited
state $|3\rangle$ through spatially varying classical laser fields, with the
Rabi frequencies $\Omega_{13}$ and $\Omega_{23}$, respectively.
In contrast to the $\Lambda$-type system, an additional coupling between
$|1\rangle$ and $|2\rangle$ is applied by the third classical laser field
with the Rabi frequency $\Omega_{12}$.

The total wave function $|\Psi(\mathbf{r})\rangle=\sum_{j=1}^{3}\psi_{j}(\mathbf{r}
)|j\rangle$ of the cyclic molecule, where $\mathbf{r}$ denotes the molecular
center-of-mass, is governed by the total Hamiltonian $H=\mathbf{p}
^{2}/(2m)+U(\mathbf{r})+H_{\mathrm{inn}}$, where $m$ is the molecular mass.
The trapping potential $U(\mathbf{r})=\sum_{j}U_{j}(\mathbf{r})|j\rangle
\langle j|$ is diagonal in the basis of inner states $|j\rangle$, and the
inner Hamiltonian $H_{\mathrm{inn}}$ contains the free terms\ $\omega
_{j}\left\vert j\right\rangle \langle j|$\ and the Rabi coupling terms
$\Omega_{jl}\exp(-i\nu_{jl}t)\left\vert j\right\rangle \langle l|+$H.c.
\ ($j=1,2,3;$ $l>j$) where $\omega_{j}$ corresponds to the inner level
energies. From now on we assume $\hbar=1$. Here, the frequencies of the three
classical optical fields are $\nu_{jl}$ matching the transition $\left\vert
j\right\rangle \rightarrow\left\vert l\right\rangle $ with the Rabi
frequencies $\Omega_{jl}=\mu_{jl}E_{jl}=\left\vert \Omega_{jl}(t)\right\vert
\exp(i\phi_{jl})$, respectively; $\mu_{jl}$ are the electrical dipole matrix
elements, and $E_{jl}$ the envelopes of electric fields corresponding to
the optical fields that couple levels $j$ and $l$; $\phi_{jl}$ are the corresponding phases.

We now consider the case that the optical field of Rabi frequency $\Omega
_{12}$ is resonant to the transition $\left\vert 1\right\rangle \rightarrow
\left\vert 2\right\rangle $, while the other two optical fields are in
two-photon resonance with the same single-photon detuning $\Delta=\omega
_{3}-\omega_{2}-\nu_{23}=\omega_{3}-\omega_{1}-\nu_{13}$ (see Fig.
\ref{Fig01}(a)). For position-independent or adiabatically varying
$\Omega_{jl}$, the inner Hamiltonian $H_{\mathrm{inn}}$ can be re-written in a
time-independent form
\begin{equation}
H_{\mathrm{inn}}^{\prime}=\Delta\left\vert 3\right\rangle \left\langle
3\right\vert +\sum_{l>j=1}^{3} \Omega_{jl}\left\vert j\right\rangle
\left\langle l\right\vert +\text{H.c.} \label{hamil}%
\end{equation}
in the interaction picture.

From now on, we assume large detuning and weak coupling: $|\Delta|\gg
|\Omega_{13}|$ $\sim|\Omega_{23}|$ $\gg|\Omega_{12}|$, so that we can use a
canonical transformation \cite{Frohlich-Nakajima,sun05} to eliminate the
excited level $\left\vert 3\right\rangle $ from the Hamiltonian (\ref{hamil}).
To this end we decompose the Hamiltonian as $H_{\mathrm{inn}}^{\prime}%
=H_{0}+H_{1}+H_{2}$ with the zeroth-order Hamiltonian $H_{0}=\Delta\left\vert
3\right\rangle \left\langle 3\right\vert $, the first-order term $H_{1}%
=\Omega_{13}\left\vert 1\right\rangle \left\langle 3\right\vert +\Omega
_{23}\left\vert 2\right\rangle \left\langle 3\right\vert +$H.c., and
second-order term $H_{2}=\Omega_{12}\left\vert 1\right\rangle \left\langle
2\right\vert +$H.c.. Then the unitary transformation
\cite{Frohlich-Nakajima,sun05} $H_{\mathrm{eff}}^{\mathrm{inn}}=\exp
(-S)H_{\mathrm{inn}}^{\prime}\exp(S)\simeq H_{0}+[H_{1},S]/2+H_{2}$ defined by
the anti-Hermitian operator $S=(\Omega_{13}\left\vert 1\right\rangle
\left\langle 3\right\vert +\Omega_{23}\left\vert 2\right\rangle \left\langle
3\right\vert -$ H.c.$)/\Delta$ results in the following second-order Hamiltonian
\begin{align}
H_{\mathrm{eff}}^{\mathrm{inn}}  &  =\Delta\left\vert 3\right\rangle
\left\langle 3\right\vert +\Lambda_{1}\left\vert 1\right\rangle \left\langle
1\right\vert +\Lambda_{2}\left\vert 2\right\rangle \left\langle 2\right\vert
\nonumber\\
&  +(ge^{i\Phi}\left\vert 1\right\rangle \left\langle 2\right\vert
+\text{H.c.}), \label{hamilL-eff}%
\end{align}
where the energy shifts $\Lambda_{i}$ are given by $\Lambda_{1}=-|\Omega
_{13}|^{2}/\Delta$, $\Lambda_{2}=-|\Omega_{23}|^{2}/\Delta$, and the effective
coupling is $g\exp(i\Phi)=\Omega_{12}-\Omega_{13}\Omega_{23}^{\ast}/\Delta$.
The instantaneous eigen-states of $H_{\mathrm{eff}}^{\mathrm{inn}}$ are
obtained as $|\chi_{3}\rangle=\left\vert 3\right\rangle $, and the dressed
states
\begin{align}
|\chi_{1}\rangle &  =\cos\theta\left\vert 1\right\rangle +e^{-i\Phi}\sin
\theta\left\vert 2\right\rangle ,\nonumber\\
|\chi_{2}\rangle &  =-\sin\theta\left\vert 1\right\rangle +e^{-i\Phi}%
\cos\theta\left\vert 2\right\rangle
\end{align}
with the corresponding eigenvalues $\lambda_{j}=\Lambda_{j}-(-1)^{j}%
g\tan\theta$ for $j=1,2$ and $\lambda_{3}=\Delta$ where $\theta$ is given by
$\tan2\theta=2g/(\Lambda_{1}-\Lambda_{2})$.

\emph{Induced gauge potentials.} In the new inner dressed-state basis
$\{|\chi_{1}\rangle,|\chi_{2}\rangle,|\chi_{3}\rangle\}$, the full quantum
state $\left\vert \Psi(\mathbf{r})\right\rangle =\sum_{j=1}^{3}\psi
_{j}(\mathbf{r})|j\rangle$ should be represented as $|\Psi(\mathbf{r}%
)\rangle=\sum_{j=1}^{3}\tilde{\psi}_{j}(\mathbf{r})|\chi_{j}\rangle$, where
the wave functions $\tilde{\psi}=(\tilde{\psi}_{1},\tilde{\psi}_{2}%
,\tilde{\psi}_{3})^{T}$ obey the Schr\"{o}dinger equation $i\partial_{t}%
\tilde{\psi}=\tilde{H}\tilde{\psi}$ with the effective Hamiltonian $\tilde
{H}=(i\mathbf{\nabla}+\mathbf{\underline{A}(r)})^{2}/(2m)+\underline
{V}(\mathbf{r})$. Here, the induced gauge potentials, i.e., the vector
potential $\mathbf{\underline{A}(r)}$ and the scalar potential $\underline
{V}(\mathbf{r})$, are two $3\times3$ matrices defined by $\mathbf{A}%
_{j,l}=i\langle\chi_{j}|\mathbf{\nabla}\chi_{l}\rangle$ and $V_{j,l}%
=\lambda_{j}\delta_{j,l}+\langle\chi_{j}|U(\mathbf{r})\mathbf{|}\chi
_{l}\rangle$. The off-diagonal elements of $\mathbf{\underline{A}}$ and
$\underline{V}$ can be neglected: the Born-Oppenheimer approximation can be
applied to show that they vanish if the adiabatic condition applies
\cite{sun-ge}. Furthermore, the inner excited state $|\chi_{3}\rangle
=\left\vert 3\right\rangle $, whose eigen-energy $\lambda_{3}=\Delta$ is much
larger than the other inner eigen-energies $\lambda_{1}$ and $\lambda_{2}$, is
decoupled from the other inner dressed states. Thus, the three-level cyclic
system is reduced to the subsystem spanned by the two lower eigenstates
$\left\{  |\chi_{1}\rangle,|\chi_{2}\rangle\right\}  $, which are robust to
atomic spontaneous emission. This results in an effective spin-1/2 system with
pseudo-spin up and down states $\left\vert \uparrow\right\rangle \equiv
|\chi_{1}\rangle$ and $\left\vert \downarrow\right\rangle \equiv|\chi
_{2}\rangle$.

The Schr\"{o}dinger equation of the effective two-level system in the
pseudo-spin-1/2 basis $\left\{  \left\vert \uparrow\right\rangle ,\left\vert
\downarrow\right\rangle \right\}  $ is governed by the diagonal effective
Hamiltonian $\tilde{H}_{\mathrm{eff}}=H_{\uparrow}\left\vert \uparrow
\right\rangle \left\langle \uparrow\right\vert +H_{\downarrow}\left\vert
\downarrow\right\rangle \left\langle \downarrow\right\vert $, where
\begin{equation}
H_{\sigma}=\frac{1}{2m}(i\mathbf{\nabla}+\mathbf{A}_{\sigma})^{2}+V_{\sigma
}(\mathbf{r}),\text{ }\left(  \sigma=\uparrow,\downarrow\right)  .
\end{equation}
Here, $\mathbf{A}_{\sigma}=i\langle\chi_{\sigma}|\mathbf{\nabla}\chi_{\sigma
}\rangle$ is the spin-dependent induced vector potential and
\begin{align}
V_{\sigma}(\mathbf{r})  &  =\lambda_{\sigma}+\langle\chi_{\sigma}%
|U|\chi_{\sigma}\rangle\nonumber\\
&  +\frac{1}{2m}[\langle\mathbf{\nabla}\chi_{\sigma}|\mathbf{\nabla}%
\chi_{\sigma}\rangle+|\langle\chi_{\sigma}|\mathbf{\nabla}\chi_{\sigma}%
\rangle|^{2}]
\end{align}
is the reduced optically-induced scalar potential \cite{Ruseckas05} for the
spin-$\sigma$ component where $\lambda_{\uparrow,\downarrow}:=\lambda_{1,2}$.

We now consider a specific configuration of three Gaussian laser beams
co-propagating in the $-\hat{z}$ direction. The spatial profiles of the
corresponding Rabi frequencies $\Omega_{jl}$ are assumed to be of Gaussian
form
\begin{equation}
\Omega_{jl}=\Omega_{jl}^{0}e^{-(x-x_{jl})^{2}/\sigma_{jl}^{2}}e^{-ik_{jl}%
z}\text{ },
\end{equation}
where $j<l=1,2,3,$ $\Omega_{jl}^{0}$ are real constants, the wave vectors
satisfy $k_{12}+k_{23}-k_{13}=0$, and the center positions are assumed to be
$x_{13}=-x_{23}=\Delta x$, $x_{12}=0$. The explicit form of the vector
potentials are
\begin{equation}
\mathbf{A}_{\uparrow}=-k_{12}\sin^{2}\theta\mathbf{\hat{e}}_{z},\text{
}\mathbf{A}_{\downarrow}=-k_{12}\cos^{2}\theta\mathbf{\hat{e}}_{z}.
\label{gauge-left}%
\end{equation}
Thus, the different spin states of the molecule will have opposite
spin-dependent effective magnetic fields $\mathbf{B}_{\uparrow}%
=-\mathbf{B_{\downarrow}}$ according to $\mathbf{B}_{\sigma}=\mathbf{\nabla
}\times\mathbf{A}_{\sigma}$ \cite{Juzeliunas2006}.

The internal state of the molecule is prepared in the spin $up$ and $down$ by
using the laser beams. The external atomic trap $U(\mathbf{r})$ is turned off
at time $t=0$, and the molecules fall due to gravity with an acceleration $G$
along the direction $\hat{z}$ \cite{note}. The scalar potentials $V_{\sigma}$
for spin-up and down molecules are given explicitly as
\begin{align}
V_{\uparrow}(\mathbf{r})  &  =\lambda_{\uparrow}+\frac{1}{2m}[k_{12}^{2}%
\sin^{2}\theta(1+\sin^{2}\theta)+\left(  \partial_{x}\theta\right)
^{2}],\nonumber\\
V_{\downarrow}(\mathbf{r})  &  =\lambda_{\downarrow}+\frac{1}{2m}[k_{12}%
^{2}\cos^{2}\theta(1+\cos^{2}\theta)+\left(  \partial_{x}\theta\right)  ^{2}].
\label{scalar-left}%
\end{align}

The spin-dependent induced vector potential $\mathbf{A}_{\sigma}(\mathbf{r})$
and scalar potential $V_{\sigma}(\mathbf{r})$ lead to the following equations
of orbital motion
\begin{align}
\dot{x}_{\sigma}  &  =\frac{p_{\sigma,x}}{m},\text{ }\dot{z}_{\sigma}%
=\frac{p_{\sigma,z}-A_{\sigma,z}}{m},\text{ }\dot{p}_{\sigma,z}=mG,\nonumber\\
\dot{p}_{\sigma,x}  &  =\frac{1}{m}\left[  (\partial_{x}A_{\sigma,z}%
)p_{\sigma,z}-A_{\sigma,z}\partial_{x}A_{\sigma,z}\right]  -\partial
_{x}V_{\sigma}.\ \label{motion equation}%
\end{align}

Hence, there will be a Stern-Gerlach-like effect, i.e., different spatial
motion of the cyclic molecules corresponding to different initial states of
spin up and down. In contrast to the standard Stern-Gerlach effect, the
effective magnetic field is not required to have a gradient. Here and in the
following, we treat the orbital motion as classical because of the large
molecular mass and weak effective gauge potentials.

\emph{Generalized Stern-Gerlach effect.} In the large detuning and
weak-coupling limit, the above approach works well for any type of cyclic
three-level system. It can also be applied in an experimentally feasible
scheme to detect the chirality of molecules, since the left- and
right-handed molecules have different Stern-Gerlach-like effects. Physically,
left- and right-handed molecules have the same intrinsic properties except the
antisymmetry of the total phase for the three coupled Rabi frequencies
\cite{Kral01,Kral03}. Hence, we can define $\Omega_{ij}^{L}\equiv\Omega_{ij}$
as the Rabi frequencies for the left-handed molecules, and define the Rabi
frequencies for the right-handed ones: $\Omega_{12}^{R}\equiv-\Omega_{12}$ and
$\Omega_{13}^{R}\equiv\Omega_{13}$, $\Omega_{23}^{R}\equiv\Omega_{23}$ for the
same coupling optical fields (see Fig. \ref{Fig01}(a,b)). Therefore the
difference in chirality leads to two different effective couplings,
\begin{equation}
g_{L/R}e^{i\Phi_{L/R}}=\pm\Omega_{12}-\frac{1}{\Delta}\Omega_{13}\Omega
_{23}^{\ast},
\end{equation}
(the first indexes of the l.h.s correspond to the above symbols of the r.h.s.)
which results in two different effective inner Hamiltonians
\begin{align}
H_{\mathrm{eff}}^{\mathrm{inn}(Q)}  &  =\Delta\left\vert 3\right\rangle
_{QQ}\left\langle 3\right\vert +\Lambda_{1}\left\vert 1\right\rangle
_{QQ}\left\langle 1\right\vert +\Lambda_{2}\left\vert 2\right\rangle
_{QQ}\left\langle 2\right\vert \nonumber\\
&  +(g_{Q}e^{i\Phi_{Q}}\left\vert 1\right\rangle _{QQ}\left\langle
2\right\vert +\text{H.c.}), \ \ (Q=L,R).
\end{align}

\begin{figure}[ptbh]
\includegraphics[bb=0 0 600 400,width=7cm]{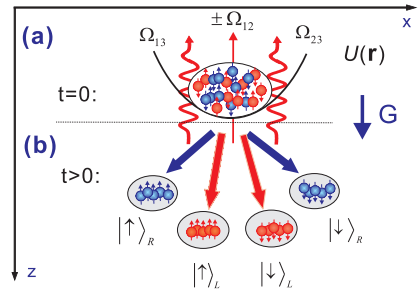}\caption{ (Color online)
Schematic illustration of the generalized Stern-Gerlach experiment of oriented  chiral molecules. (a) Mixed chiral molecules trapped by the external potential
$U(\mathbf{r})$, are coupled to three laser fields and reduce to the lower
dressed internal state-space of spin-up and spin-down. (b) After the trap
potential is switched off at time $t=0$, the molecules will fall under gravity
$G=9.8$ m/s$^{\text{2}}$. Due to the chirality-dependent induced gauge field,
molecules with different chirality will experience different Stern-Gerlach
effects.}%
\label{Fig02}%
\end{figure}

Initially, the mixed left- and right-handed oriented molecules, which are
spatially confined due to the external trap potential, 
are subject to the three coupling optical fields as seen in Fig. \ref{Fig02}(a) and reduced to the spin-state space $\{\left\vert \uparrow\right\rangle
_{L/R},\left\vert \downarrow \right\rangle _{L/R}\}$. At time $t=0$,
the external trap potential is turned off and the molecules will fall
due to gravity. As in the above consideration for the general case of
cyclic-type molecules in 
Eqs.~(\ref{hamilL-eff})-(\ref{motion equation}), 
we can obtain the optically-induced potentials and
molecular classical trajectories for left- and right-handed molecules,
respectively. This is schematically illustrated in
Fig. \ref{Fig02}(b), which shows that the generalized Stern-Gerlach
effect splits the initial cloud into four subsets, since the effective
gauge potentials depend both on spin and chirality.

\begin{figure}[h]
\includegraphics[bb=0 20 320 100, width=8cm, clip]{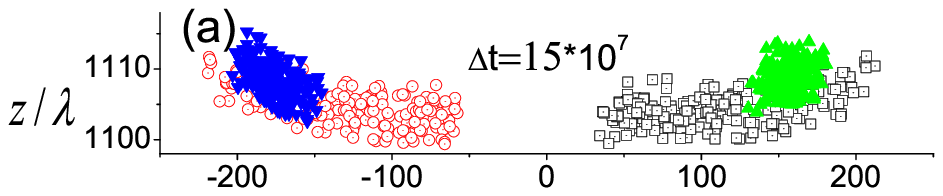}
\includegraphics[bb=0 20 320 100, width=8cm, clip]{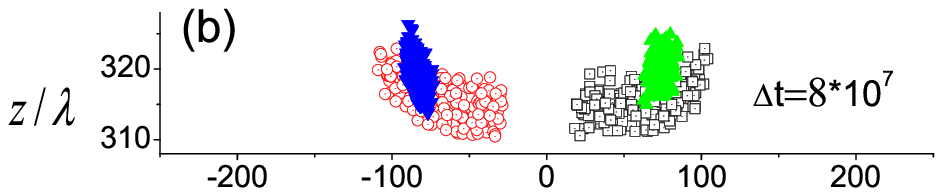}
\includegraphics[bb=0 20 320 100, width=8cm, clip]{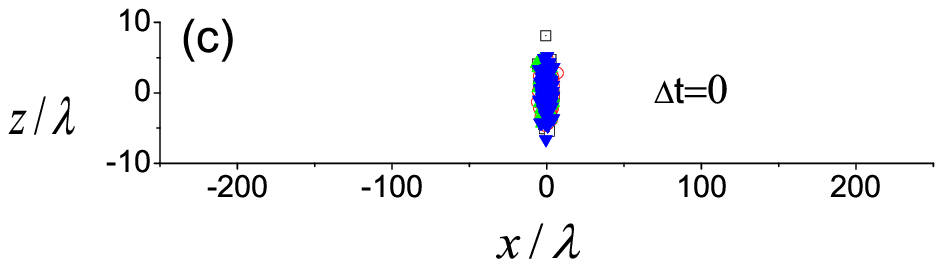}
\includegraphics[bb=0 20 320 100, width=8cm, clip]{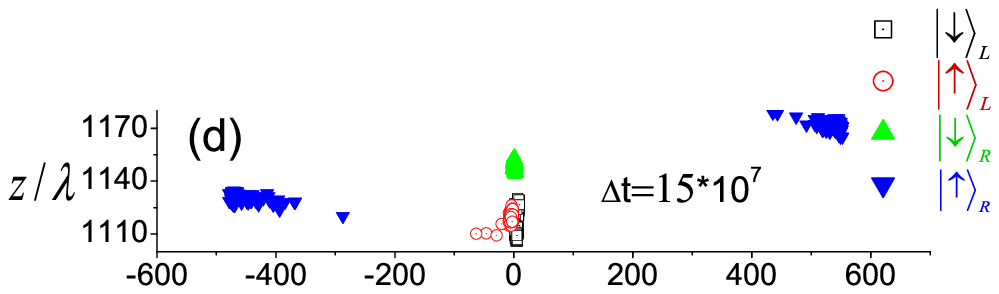}
\includegraphics[bb=0 20 320 100, width=8cm, clip]{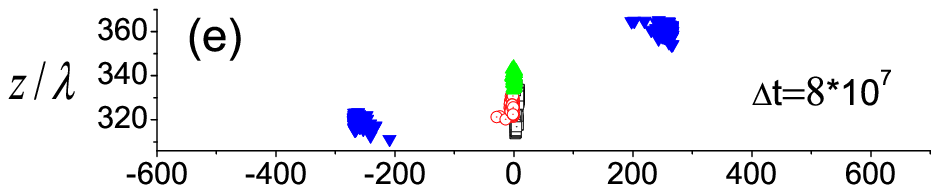}
\includegraphics[bb=0 20 320 100, width=8cm, clip]{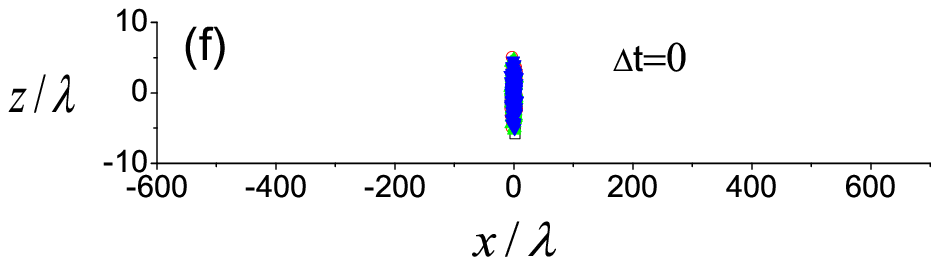}
\caption{(Color online) The positions of an oriented molecular
ensemble with an initial Gaussian position distribution
($\rho(x,z)$=$(2\pi\sigma_{r}^{2})^{-1}$ $\exp
[-(x^{2}+z^{2})/\sigma_{r}^{2}]$ with $\sigma_{r}=3\lambda$) at
different times. The positions $x,z$ are in units of $\lambda$ (the
wavelength for the lower transition $\lambda=2\pi/k_{12}=2\pi
c/\nu_{12}$ with $c$ the optical velocity in vacuum; typically,
$\lambda\sim1$ $\mu$m). The following parameters are used: (a-c):
$\Omega_{12}^{0}=\Omega_{13}^{0}\Omega_{23}^{0}/\Delta\sim10^{-6}
\Delta$; (d-f): $\Omega_{12}^{0}=\Omega_{13}^{0}
\Omega_{23}^{0}/\Delta\sim10^{-4} \Delta$ (the detuning
$\Delta\sim10^{10}$ Hz is large). Here, we assume
$\sigma_{13}\equiv\sigma_{23}=10\lambda$, $\sigma_{12}=7\lambda$,
$\Delta x=3\lambda$, $\Omega_{13}^{0}\equiv\Omega _{23}^{0}$. The
molecular mass is taken to be 100 times the proton mass.}%
\label{Fig03}%
\end{figure}

\begin{figure}[h]
\includegraphics[height=3 cm,width=8 cm]{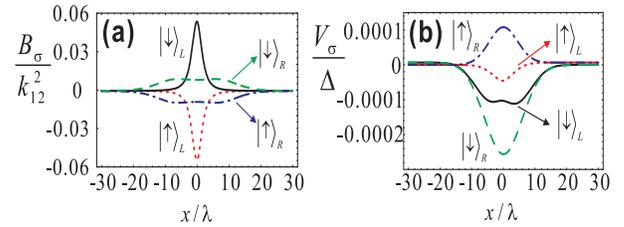}\caption{(Color online)
(a) Effective magnetic field corresponding to Fig. \ref{Fig03} (in units of
$k_{12}^{2}$ $\equiv$ $4\pi^{2}/\lambda^{2}$); (b) Effective scalar potential
corresponding to Fig. \ref{Fig03}(d-f).}%
\label{Fig04}%
\end{figure}

To make this picture of a generalized Stern-Gerlach effect more quantitative,
we show in Fig.~\ref{Fig03} the typical position of an oriented 
ensemble of mixed left-
and right-handed molecules and spin states subject to gravity (in the $\hat
{z}$-direction). For temperatures below $1$ $\mu$K, the initial velocity of
the molecules can be neglected. Figures~\ref{Fig03}(a-c) show the $\hat{x}%
$-$\hat{z}$-plane positions of such a molecular ensemble with an initial
Gaussian position distribution at the origin at different times. The spatial
separation of molecules with different spin projections is clearly visible. By
choosing a different value of $\Omega_{12}^{0}$ and $\Omega_{13}^{0}%
\Omega_{23}^{0}/\Delta$, we also obtain a spatial separation of molecules with
different chirality, see Fig. \ref{Fig03}(d-f). The separation is partial in
the following sense: for our choice of parameters, right-handed molecules in
the spin-up state are deflected to finite values of $x$, whereas the other
three components are not deflected and their trajectories remain close to
$x=0$. By changing $\Omega_{12}^{0}\rightarrow-\Omega_{12}^{0}$, the role of
left and right in Fig. \ref{Fig03}(a-f) is interchanged. In Fig. \ref{Fig04}
we show the effective magnetic fields (i.e., the curl of the vector
potentials) and scalar potentials leading to this behavior. Figure
\ref{Fig04}(a) shows the effective magnetic fields corresponding to all the
subplots in Fig. \ref{Fig03} [the effective magnetic field is the same in
Figs. 3(a-c) and 3(d-f)]. For the situation in Fig. \ref{Fig03}(a-c), the
effects of the scalar potential (which is not shown) can be neglected: the
magnetic fields are dominant and make the molecules in the spin-up state move
along the $-x$-direction (spin-down states along the $x$-direction). In
Fig.~\ref{Fig03}(d-f) the scalar potentials are dominant (Fig.~\ref{Fig04}(b))
and will trap the molecules in the area around $x=0$, except for the
$\left\vert \uparrow\right\rangle _{R}$ molecules that are deflected.

Stern-Gerlach experiments can also be used to obtain and measure
superpositions of spin states. 
However, our effect described above does not work for
superpositions of left- and right-handed chiral states (even if many studies about teleporting, preparating and measuring superpositions of chiral states \cite{harris} appeared recently), since this
would require considering higher excited symmetric/antisymmetric
states.  We will leave this interesting question for future works.

Although the protocol presented here is idealized since inter-molecular
interactions are neglected, it provides a promising way to spatially separate
molecules of different chiralities. A similar generalized Stern-Gerlach
effect has been proposed for $\Lambda$-type systems where the Rabi
frequencies $\Omega_{12}$ between the two lower inner states vanish
\cite{zhu}. However, this effect is chirality-independent. Thus, in contrast
to our configuration, the effect discussed in \cite{zhu} cannot be used to
distinguish and separate left- and right-handed molecules.

\emph{Conclusion.} In conclusion, we have studied the orbital effects
of internal adiabatic transitions on the center-of-mass motion of oriented 
chiral molecules. We have shown that under the conditions described above,
the center-of-mass motion of the molecules depends on both chirality
and spin due to the optically induced gauge potentials and can be
interpreted as a generalized Stern-Gerlach effect. This leads to the
possibility of spatially separating molecules of different
chiralities.

This work was supported by the European Union under contract
IST-3-015708-IP EuroSQIP, by the Swiss NSF, and the NCCR Nanoscience,
and also by the NSFC and NFRPC of China.


\begin{thebibliography}{99}

\bibitem {SG}D. Bohm, \textit{Quantum Theory} (Prentice-Hall, Englewood
Cliffs, NJ, 1951).

\bibitem {wilz}F. Wilczek and A. Zee, Phys. Rev. Lett. \textbf{52},
2111 (1984).

\bibitem {moody}J. Moody \textit{et al.},
Phys. Rev. Lett. \textbf{56}, 893 (1986).

\bibitem {sun-ge}C. P. Sun and M. L. Ge, 
Phys. Rev. D \textbf{41}, 1349 (1990).

\bibitem {zhang-li-sun}P. Zhang \textit{et al.},
Eur. Phys. J. D \textbf{36}, 229 (2005).

\bibitem {zhu}S.-L. Zhu \textit{et al.},
Phys. Rev. Lett. \textbf{97}, 240401 (2006).

\bibitem {liu}X.-J. Liu \textit{et al.},
Phys. Rev. Lett. \textbf{98}, 026602 (2007).

\bibitem {SHE}J. E. Hirsch, Phys. Rev. Lett. \textbf{83}, 1834 (1999); 
S. Murakami \textit{et al.},
Science \textbf{301}, 1348 (2003); S. Murakami \textit{et al.},
Phys. Rev. Lett. \textbf{93}, 156804 (2004); J. Sinova \textit{et al.},
Phys. Rev. Lett. \textbf{92,} 126603 (2004); Y. K. Kato \textit{et al.},
Science \textbf{306}, 1910 (2004).

\bibitem {Kral01}P. Kr\'{a}l and M. Shapiro, 
Phys. Rev. Lett. \textbf{87}, 183002 (2001).

\bibitem {Kral03}P. Kr\'{a}l \textit{et al.},
Phys. Rev. Lett. \textbf{90}, 033001 (2003).

\bibitem {shapiro}M. Shapiro, E. Frishman, and P. Brumer, 
Phys. Rev. Lett. \textbf{84}, 1669 (2000).


\bibitem {Liu05}Y.-X. Liu \textit{et al.},
Phys. Rev. Lett. \textbf{95}, 087001 (2005).

\bibitem {cyclic}P. Kr\'{a}l \textit{et al.},
Phys. Rev. A \textbf{72}, 020303(R) (2005); Y. Li \textit{et al.},
Phys. Rev. A \textbf{73}, 043805 (2006); J. Q. You \textit{et al.},
quant-ph/0512145.

\bibitem {sun05}C. P. Sun \textit{et al.},
quant-ph/0506011.

\bibitem {Frohlich-Nakajima}H. Fr\"{o}hlich, Phys. Rev. \textbf{79}, 845
(1950); S. Nakajima, Adv. Phys. \textbf{4}, 363 (1955).

\bibitem {Ruseckas05}J. Ruseckas \textit{et al.},
Phys. Rev. Lett. \textbf{95}, 010404 (2005).

\bibitem {Juzeliunas2006}G. Juzeliunas and P. Ohberg,
Phys. Rev. Lett. \textbf{93}, 033602 (2004); G. Juzeliunas \textit{et al.},
Phys. Rev. A \textbf{73}, 025602 (2006).

\bibitem {note}Alternatively, instead of the gravitational field, we could use
an initial velocity of the molecules in the direction $\hat{z}$.

\bibitem {harris}C. S. Maierle and R. A. Harris, 
J. Chem. Phys. \textbf{109}, 3713 (1998); 
C. S. Maierle, D. A. Lidar, and R. A. Harris, 
Phys. Rev. Lett. \textbf{81}, 5928 (1998);  
R. A. Harris and J. D. Walls, 
AIP Conference Proceedings \textbf{596}, 186 (2001).

\end{thebibliography}
\end{document}